\documentclass[aps,prl,reprint,twocolumn,superscriptaddress,longbibliography,nofootinbib,floatfix,showpacs]{revtex4-1}
\usepackage{epsfig}
\usepackage{amsmath,amssymb,amsfonts}
\usepackage{mathrsfs}
\usepackage{bbm}
\usepackage{slashed}
\usepackage{graphicx}
\usepackage{verbatim}

\usepackage[T1]{fontenc}
\usepackage[latin1]{inputenc}
\usepackage[colorlinks]{hyperref}

\usepackage{ifthen}
\usepackage{forloop}
\usepackage[english]{babel}
\usepackage{mathbbol}

\usepackage[usenames]{color}

\definecolor{darkgreen}{rgb}{0.2,0.6,0}
\definecolor{lightblue}{rgb}{0,0.5,0.8}
\definecolor{lightred}{rgb}{0.8,0.2,0.2}
\definecolor{darkorange}{rgb}{1,0.549,0}

\newcommand{\Eqref}[1]{Eq.~\eqref{#1}}
\newcommand{\be}{\begin{equation}}
\newcommand{\ee}{\end{equation}}
\newcommand{\bw}{\begin{widetext}}
\newcommand{\ew}{\end{widetext}}
\newcommand{\bi}{\begin{itemize}}
\newcommand{\ei}{\end{itemize}}
\newcommand{\bea}{\begin{eqnarray}}
\newcommand{\eea}{\end{eqnarray}}

\newcommand{\cR}{\mathcal R}
\newcommand{\p}{\partial}

\newcommand{\gb}{\bar{g}}
\newcommand{\gs}{\sigma}

\newcommand{\UV}{{\small UV}}
\newcommand{\IR}{{\small IR}}

\newcommand{\RG}{{\small RG}}
\newcommand{\GFP}{{\small GFP}}
\newcommand{\NGFP}{{\small NGFP}}
\newcommand{\CPU}{{\small CPU}}
\newcommand{\eg}{{\textit{e.g.}}}

\begin{document}

\title{The Gravitational Two-Loop Counterterm is Asymptotically Safe}

\author{Holger Gies}
\email[]{holger.gies@uni-jena.de}
\affiliation{Theoretisch-Physikalisches Institut, Abbe Center of Photonics, Friedrich-Schiller-Universit\"at Jena, 
Max-Wien-Platz 1, D-07743 Jena, Germany}   
%\affiliation{Helmholtz-Institut Jena, Fr\"obelstieg 3, D-07743 Jena, Germany}
\author{Benjamin Knorr}
\email[]{benjamin.knorr@uni-jena.de}
\affiliation{Theoretisch-Physikalisches Institut, Abbe Center of Photonics, Friedrich-Schiller-Universit\"at Jena, 
Max-Wien-Platz 1, D-07743 Jena, Germany}   
\author{Stefan Lippoldt}
\email[]{stefan.lippoldt@uni-jena.de}
\affiliation{Theoretisch-Physikalisches Institut, Abbe Center of Photonics, Friedrich-Schiller-Universit\"at Jena, 
Max-Wien-Platz 1, D-07743 Jena, Germany}   
\author{Frank Saueressig}
\email[]{f.saueressig@science.ru.nl}
\affiliation{
Institute for Mathematics, Astrophysics and Particle Physics (IMAPP),\\
Radboud University Nijmegen, Heyendaalseweg 135, 6525 AJ Nijmegen, The Netherlands
}

\begin{abstract}
Weinberg's asymptotic safety scenario provides an elegant mechanism to
construct a quantum theory of gravity within the framework of quantum
field theory based on a non-Gau{\ss}ian fixed point of the
renormalization group flow. In this work we report novel evidence for
the validity of this scenario, using functional renormalization group
techniques to determine the renormalization group flow of the
Einstein-Hilbert action supplemented by the two-loop counterterm found
by Goroff and Sagnotti. The resulting system of beta functions
comprises three scale-dependent coupling constants and exhibits a
non-Gau{\ss}ian fixed point which constitutes the natural extension of
the one found at the level of the Einstein-Hilbert action. The fixed
point exhibits two ultraviolet attractive and one repulsive direction
supporting a low-dimensional \UV{}-critical hypersurface. Our result
vanquishes the longstanding criticism that asymptotic safety will not
survive once a ``proper perturbative counterterm'' is included in the
projection space.
\end{abstract}

\maketitle

\section{Introduction}
General relativity, based on the Einstein-Hilbert action, provides a
highly successful classical description of gravitational phenomena
from sub-millimeter to cosmic scales. A central puzzle for the
construction of a consistent quantum theory of gravity is its
perturbative nonrenormalizability. This is manifested by the fact,
that an expansion in terms of
Newton's constant about flat spacetime
gives rise to a divergence at two-loop order. This spoils meaningful
predictions for $S$-matrix elements, unless a Goroff-Sagnotti
counter-term of the form
\cite{Goroff:1985sz,Goroff:1985th,vandeVen:1991gw}
\begin{equation}
\Gamma_{\text{GS}}= \frac{1}{\epsilon} \frac{209}{2880} \frac{1}{(16\pi^2)^2} \int d^4x
\sqrt{g} \,  C_{\mu\nu}{}^{\kappa\lambda}C_{\kappa\lambda}{}^{\rho\sigma}C_{\rho\sigma}{}^{\mu\nu}
\label{eq:GS}
\end{equation}
with 
the Weyl-tensor $C_{\mu\nu\rho\sigma}$, is added to the
bare action in order to cancel the divergence (in dimensional
regularization). In combination with power-counting arguments, this is
taken as a signal that an infinite number of counterterms is needed for
rendering the full perturbative expansion meaningful.  Since
renormalization theory relates each counterterm to a free parameter
to be fixed from experimental data, the appearance of the
Goroff-Sagnotti term suggests that the perturbative quantization of
the Einstein-Hilbert action requires fixing infinitely many
parameters.  This observation is often interpreted as evidence that
conventional quantization of gravity is doomed to fail.

The presence of the counterterm \eqref{eq:GS} triggered the
investigation of a variety of alternative routes towards quantizing
gravity, e.g., by modifying the quantization rules, changing the
fundamental degrees of freedom, or abandoning local quantum field
theory altogether as a fundamental framework for quantum gravity
\cite{Ashtekar:2005bf,Hamber:2009zz,Kiefer:2012boa}. Ultimately, any
consistent quantum gravity theory allowing for a classical limit
containing Einstein's theory of gravity,
as well as its semi-classical extension as a low-energy effective
theory for quantized gravitons, has to clarify the fate of the
divergencies related to the Goroff-Sagnotti term.

This requirement is more than a technical necessity, as
renormalizability -- beyond being a strategy of handling divergencies
-- is a statement about the separability of low-energy observables
from physics at highest energy scales. For instance, quantum gravity
scenarios that start from discretized building blocks of spacetime and
a fundamental length scale may render all divergencies finite. Still,
$S$-matrix elements would generically receive large contributions from
potentially large higher-order operators, requiring to fix a
substantial, if not infinite, set of physical parameters.

At first sight, a natural solution appears to be that the divergencies
cancel, \eg, because of a new symmetry at a more fundamental level.
While certainly possible, the problem of separation of low-energy
physics from highest energy scales may come in again through the
backdoor, as this symmetry has to be broken (or restored) at low
energy potentially requiring a fine-tuned separation of scales and a
large number of parameters.

An indiscriminate association of \eqref{eq:GS} with quantum field
theory approaches to quantum gravity ignores the fact that the
Wilsonian viewpoint of renormalization already offers a solution to
this puzzle: higher-dimensional operators decouple from the low-energy
physics proportional to an inverse power of a high scale $\Lambda$,
provided such operators do not acquire
large anomalous dimensions. For instance, the $C^3$ operator in \eqref{eq:GS}
would be expected to decouple $\sim 1/\Lambda^2$ if the anomalous
dimension was small. If so, the $1/\epsilon$-pole may merely indicate
a subleading log-correction as sensed by dimensional regularization.
This may sound like a circular argument, as such conclusions can only
be drawn in perturbation theory after the theory has been
renormalized. Nevertheless, the Wilsonian viewpoint is
known to hold also in systems with a similar breakdown of perturbative
quantization, where a well-controllable ultraviolet limit is
facilitated by the existence of an interacting renormalization group
(\RG{}) fixed point
\cite{Wilson:1972cf,Weinberg:1976xy,Weinberg:1980gg}. 

In the present letter, we provide novel evidence that the
Goroff-Sagnotti term is indeed an irrelevant operator from the
Wilsonian viewpoint. Our results demonstrate that the challenge posed
by the perturbative two-loop analysis is solved by a renormalization
flow that decouples the high-scale physics from 
(semi-)classical Einsteinian low-energy gravity 
in much the same way as in conventional quantum field theories. For
this, we determine the decoupling of the Goroff-Sagnotti term towards
the \IR{} quantitatively.

The 
new ingredient compared to the perturbative analysis is the
investigation of the \RG{} flow beyond the perturbative Gau\ss{}ian
fixed point (\GFP{}). In fact, our results confirm the existence of an
interacting non-Gau\ss{}ian fixed point (\NGFP{}) that controls the
high-energy limit of gravity, as required for the asymptotic safety
scenario
\cite{Weinberg:1976xy,Weinberg:1980gg,Reuter:1996cp,Niedermaier:2006wt,Codello:2008vh,Litim:2011cp,Reuter:2012id}.
By now, the existence of a suitable \NGFP{} has been established
within many approximations
\cite{Lauscher:2001rz,Lauscher:2002sq,Codello:2006in,Codello:2007bd,Machado:2007ea,%
  Bonanno:2010bt,Falls:2013bv,Benedetti:2009rx,Benedetti:2009gn,%
  Rechenberger:2012pm,Demmel:2015oqa,Eichhorn:2009ah,Groh:2010ta,Eichhorn:2010tb,%
  Christiansen:2015rva,Nink:2012vd,Christiansen:2012rx,%
  Christiansen:2014raa,Becker:2014jua,Becker:2014qya,Becker:2014pea}. In
particular, it has been shown in the case of gravity coupled to scalar
matter that the asymptotic safety mechanism remains intact once the
one-loop counterterm is included
\cite{Benedetti:2009rx,Benedetti:2009gn}. Paralleling this
observation, we establish that the Goroff-Sagnotti term supplements
only a subdominant quantitative correction to the high-energy behavior
of pure gravity: the $C^3$ operator approaches an interacting
fixed point in the \UV{} and becomes irrelevant towards the \IR{} at
an even enhanced rate compared to canonical scaling.

This demonstrates that the asymptotic safety scenario for quantum
gravity can solve this long-standing puzzle in a constructive and
quantifiable manner, disclosing the two-loop divergence of
\eqref{eq:GS} as a mere perturbative artifact.

%--------------------------------------------------------
\section{Functional renormalization}
%--------------------------------------------------------
A powerful tool to investigate 
renormalizability
based on interacting \RG{} fixed points is provided by the
functional \RG{}. In the incarnation based on the effective average action
$\Gamma_k$ \cite{Wetterich:1992yh}, the functional \RG{} equation 
\be\label{FRGE}
k\p_k \Gamma_k = \tfrac{1}{2} {\rm Str} \left[ \left( \Gamma_k^{(2)} + \cR_k \right)^{-1} \, k \p_k \cR_k \right] 
\ee
realizes Wilson's idea of
renormalization by integrating out quantum fluctuations shell-by-shell in momentum space. The use of the two-point correlator $\Gamma_k^{(2)}$ leads
to a formally exact equation. Owed to the regulator $\cR_k$ the change
of $\Gamma_k$ is driven by 
quantum fluctuations with
momenta close to $k$. A key advantage of the functional \RG{} is that it
permits approximations which do not rely on 
a small
expansion parameter. Moreover, the flow equation allows investigating
\RG{} flows without the need of specifying a
fundamental action a priori. This makes the setup predestined for
searching for fixed points of the 
renormalization flow
beyond the realm of perturbation theory.

We focus on the case where the gravitational degrees of freedom are carried by the spacetime metric $g_{\mu\nu}$. 
The flow equation \eqref{FRGE} can then be constructed via the background-field method splitting the full metric in a background metric $\bar{g}_{\mu\nu} $ and
fluctuations $h_{\mu\nu}$, $g_{\mu\nu} = \bar{g}_{\mu\nu} + h_{\mu\nu}$ \cite{Reuter:1996cp}. In this way the effective average action can 
be computed in a covariant way.

\section{Projection of the flow equation}
We study the gravitational renormalization flow projected on the
Einstein-Hilbert action supplemented by the two-loop counterterm \eqref{eq:GS}.
Our ansatz for the gravitational part of the effective average action, closely following
\cite{vandeVen:1991gw}, 
reads
\be\label{ansatz}
\Gamma_k = \Gamma_k^{\rm EH} + \Gamma_k^{\rm GS} 
\, . 
\ee
Here
\be\label{EHaction}
\Gamma_k^{\rm EH} = \frac{1}{16 \pi G_k} \int d^4x \sqrt{g} \left(-R + 2 \Lambda_k \right)
\ee
is the Einstein-Hilbert action including a scale-dependent Newton's constant and cosmological constant, and
\be\label{GSaction}
\Gamma_k^{\rm GS} = \bar \sigma_k \int d^4x \sqrt{g} \, C_{\alpha\beta}{}^{\mu\nu} C_{\mu\nu}{}^{\rho\sigma} C_{\rho\sigma}{}^{\alpha\beta}
\ee
is the two-loop counterterm found by Goroff and Sagnotti with a
scale-dependent coupling
$\bar \sigma_k$. The gravitational
part of the effective average action is supplemented by a standard
gauge-fixing procedure and we adhere to the harmonic gauge used in
\cite{Reuter:1996cp}.  The perturbative result \eqref{eq:GS} suggests
that $\bar{\sigma}_k$ diverges at least as $\ln k$ for $k\to\infty$
even in the flat-space on-shell limit $\Lambda_k\to0$ and after the
Newton coupling has been renormalized.

The \RG{} flow of the 
couplings is found by
substituting the ansatz \eqref{ansatz} into \Eqref{FRGE} and computing
the coefficients multiplying the curvature terms appearing in
Eqs.~\eqref{EHaction} and \eqref{GSaction}.  The evaluation of the
trace utilizes the technology of the universal \RG{} machine
\cite{Benedetti:2010nr} together with off-diagonal heat-kernel methods
\cite{Barvinsky:1985an,Decanini:2005gt,Anselmi:2007eq,Groh:2011vn,Groh:2011dw}.

Two crucial features make this formidable computation feasible:
firstly, we use the Ricci scalar, Ricci tensor, and Weyl tensor to
construct a basis for the interaction monomials containing a fixed
number of covariant derivatives.  The second functional derivative of
\eqref{GSaction} results in a sum of terms containing at least one
power of the Weyl tensor.  Since $C$ is trace-free by construction,
all its contractions with the metric vanish.  This entails that there
is no feedback of the Goroff-Sagnotti term on the renormalization flow
of Newton's constant and the cosmological constant. We conclude
already at this point that the asymptotic safety properties observed
in the Einstein-Hilbert sector are stable upon the inclusion of the
Goroff-Sagnotti term. Secondly, the contribution of the
Goroff-Sagnotti term to the two-point correlator is of the form
$\bar{\sigma}_k \left(C + \mbox{higher powers of the curvature}
  \right)$.  This structure implies that the $\beta$ function encoding
  the flow of $\bar{\sigma}_k$ is a cubic in $\bar{\sigma}_k$ with
coefficients depending on Newton's constant and the
cosmological constant.  As a cubic has at least one zero, also the
Goroff-Sagnotti coupling must have a fixed point and hence the
associated dimensionless coupling does not necessarily diverge for
$k\to\infty$. The remaining crucial question is whether the $C^3$ term
is a relevant (as suggested by perturbation theory) or an irrelevant
operator. In case of irrelevance, the Goroff-Sagnotti term does
neither require the fixing of an additional physical parameter nor
induces a proliferation of 
counterterms.

In order to determine the coefficients of 
this cubic it
suffices to isolate the term 
$\sim C^3$ from the trace of
Eq.\ \eqref{FRGE}. As the curvature terms are orthogonal,
any term containing a Ricci scalar or Ricci tensor will not contribute
to $C^3$ and it is sufficient to keep track of powers of the Weyl
tensor and its covariant derivatives.  Formally, this can be achieved
with a
background metric $\gb_{\mu\nu}$ 
of a
Ricci-flat $K3$-surface.  Terms contributing to the basis monomial
Eq.\ \eqref{GSaction} appear in three different tensor structures
which are related by
\be
\begin{split}
	C^{\alpha}{}_{\mu}{}^{\beta}{}_{\nu} C^\mu{}_\rho{}^\nu{}_\sigma C^{\rho}{}_\alpha{}^\sigma{}_\beta = & \, \frac{1}{2}  C_{\mu\nu}{}^{\rho\sigma} C_{\rho \sigma}{}^{\alpha\beta} C_{\alpha\beta}{}^{\mu\nu} \, , \\
	C_{\mu\nu\rho\sigma} D^2 C^{\mu\nu\rho\sigma} \simeq & \, - 3 C_{\mu\nu}{}^{\rho\sigma} C_{\rho\sigma}{}^{\alpha\beta} C_{\alpha\beta}{}^{\mu\nu} \, ,
\end{split}
\ee 
where $\simeq$ denotes that the identity holds up to terms containing
the Ricci scalar and Ricci tensor. The vertices entering the
computation have been constructed with the Mathematica package xAct
\cite{xActwebpage, 2007CoPhC.177..640M,
  2008CoPhC.179..586M,2008CoPhC.179..597M, Brizuela:2008ra,
  2014CoPhC.185.1719N}. Employing the simplifications of a
$K3$-background, the Goroff-Sagnotti vertex contains 900 terms whereas
the Einstein-Hilbert vertex has only one term. The computation
was done with xAct within one month of \CPU{} time on a core with
2.8 GHz. Most of the \CPU{} time is used for the two vertex diagram due
to the enormous number of terms generated by the product rule for 
covariant derivatives. This makes the present computation quite
formidable and complementary to the recent progress in vertex
expansions of quantum gravity on flat spacetime
\cite{Christiansen:2015rva} 
of similar complexity.

\section{\texorpdfstring{$\beta$}{beta} functions}
The \RG{} flow resulting from the ansatz \eqref{ansatz} is conveniently written in terms of dimensionless couplings
$g_i \equiv \{\lambda \, , \,  g \, , \, \gs\}$, 
\be
\lambda \equiv \Lambda_k \, k^{-2} \, , \; \; \; g \equiv G_k \, k^2 \, , \; \; \; \gs \equiv \bar{\sigma}_k k^{2} \, , 
\ee
and expressed in terms of the $\beta$ functions
\be
k \partial_k \, g_{i} \equiv \beta_{g_i}(\lambda, g, \gs) \, . 
\ee
The $\beta$ functions for the dimensionless Newton's constant and
cosmological constant have been known since the beginning of the
asymptotic safety program \cite{Reuter:1996cp}. In four spacetime
dimensions and for the Litim regulator \cite{Litim:2001up} they read
\be\label{betaeh}
\begin{split}
\beta_g = & \, \left( 2 + \eta_N \right) \, g \, , \\
\beta_\lambda = & \, \left( \eta_N - 2 \right) \, \lambda + \tfrac{g}{2\pi}
\left( \tfrac{5}{1-2\lambda} - 4 - \tfrac{5}{6} \eta_N \tfrac{1}{1-2\lambda} \right).
\end{split}
\ee
Here $\eta_N$ denotes the anomalous dimension of Newton's constant,
\be
\eta_N = \frac{g \, B_1}{1-g B_2} \, ,
\ee
with
\be
\begin{split}
B_1 = & \frac{1}{3\pi} \left(  \tfrac{5}{1-2\lambda} - \tfrac{9}{(1-2\lambda)^2} - 5 \right) \, , \\
B_2 = & - \frac{1}{6\pi} \left( \tfrac{5}{2} \tfrac{1}{(1-2\lambda)} -  \tfrac{3}{(1-2\lambda)^2}  \right) \, .
\end{split}
\ee
The ansatz \eqref{ansatz} complements this system by a $\beta$
function for $\sigma$,
\be\label{victory}
\beta_\sigma = c_0 + \left(2 + c_1 \right) \sigma + c_2 \, \sigma^2 + c_3 \, \sigma^3 \, ,
\ee 
where the coefficients $c_i(g,\lambda)$ are given by
\begin{eqnarray}
	c_0 &= & \, \tfrac{1}{64\pi^2 (1-2\lambda)} \left( \tfrac{2 - \eta_N}{2 (1 - 2 \lambda )}  + \tfrac{6 - \eta_N}{(1 - 2 \lambda )^3} - \tfrac{5 \eta_N}{ 378 } \right) \, , \nonumber\\
	c_1 &= & \, \tfrac{3g}{16\pi(1-2\lambda)^2} \left( 5(6-\eta_N) + \tfrac{23(8-\eta_N)}{8(1-2\lambda)} - \tfrac{7(10-\eta_N)}{10(1-2\lambda)^2} \right) \, , \nonumber\\
	c_2 &= & \, \tfrac{g^2}{2(1-2\lambda)^3} \left( \tfrac{233(12-\eta_N)}{10} - \tfrac{9(14-\eta_N)}{7(1-2\lambda)} \right) \, , \nonumber\\ 
	c_3 &= & \,  \tfrac{6 \pi g^3 (18-\eta_N)}{(1-2\lambda)^4} \, . 
\label{eq:ccoeffs}
%\nonumber
\end{eqnarray}
We emphasize that the highest-order coefficient $c_3$ is positive for
any admissible $\lambda$, positive Newton coupling $g>0$, and
$\eta_N<18$. We have verified that $c_3$ is gauge independent, and
that its positivity is independent of the metric parametrization
\cite{Nink:2014yya,Demmel:2015zfa,Percacci:2015wwa,Falls:2015cta,%
  Falls:2015qga,Gies:2015tca,Labus:2015ska}. The $\beta$ function
\Eqref{victory} is computed {\it for the first time} and constitutes
the main result of this work.

\section{Fixed points and RG flow}
The Wilsonian viewpoint links 
renormalizability 
to
fixed points $g_{i,*}$ of the underlying \RG{} flow where
$\beta_{g_i}|_{g_{j,*}} = 0$. Linearizing the $\beta$ functions at a
fixed point, local properties of the 
flow are encoded in the
stability coefficients $\theta$ defined as minus the eigenvalues of
the stability matrix $B_{ij} \equiv \p_{g_j}
\beta_{g_i}|_{g_*}$. Relevant directions, corresponding to free
parameters of the theory 
to be fixed by experiment, are
associated with stability coefficients with a positive real part.

Already the first 
calculations
\cite{Reuter:1996cp,Souma:1999at,Reuter:2001ag} revealed that the
system \eqref{betaeh} exhibits a \GFP{} and an \NGFP{}
\be\label{fpstructure}
\begin{array}{lll}
\mbox{\GFP{}}^{\rm EH}: \; & \lambda_* =0, & g_* = 0  \\
\mbox{\NGFP{}}^{\rm EH}: \; & \lambda_* = 0.193, & g_* =0.707 \, . 
\end{array} 
\ee
The \GFP{} corresponds to a free 
theory and is a
saddle-point: trajectories with a positive Newton coupling do not end
at the \GFP{} at high energies, reflecting
the perturbative
non-renormalizability of the Einstein-Hilbert action in the Wilsonian
framework. The \NGFP{} exhibits a complex pair of stability
coefficients
\be\label{stabcoeff} \theta_{1,2} = 1.475 \pm 3.043 \, i
\, .  
\ee 
Thus the \NGFP{} is \UV{} attractive for both Newton's constant and
the cosmological constant making it suitable for asymptotic safety.

The $\beta$ function \eqref{victory} clarifies
the fate of
the fixed point structure \eqref{fpstructure} once the counterterm
\eqref{GSaction} is taken into account. Substituting $\lambda_* = g_*
= 0$ into the $\beta$ function for $\sigma$ shows that the \GFP{}$^{\rm
  EH}$ is mapped to
\be\label{gfpgs}
\begin{array}{llll}
	\mbox{\GFP{}}^{\rm GS}: \; & \lambda_* =0, & g_* = 0, & 
	\sigma_* = - \tfrac{7}{128 \pi^2}.
\end{array} 
\ee 
The stability coefficients of this fixed point coincide with the
classical mass dimension of the coupling constants;
the \GFP{}
remains a saddle point.
	
Focusing on the \NGFP{}, it is illuminating to first study the
Einstein-Hilbert induced approximation of the $\beta$ function, where
only the terms originating from \eqref{EHaction} contribute to the
running of $\sigma$. Since the contribution of the counterterm to the
$\beta$ function \eqref{victory} is captured by the coefficients $c_1,
c_2$, and $c_3$ this approximation corresponds to setting $c_1 = c_2 =
c_3 = 0$. In this limit the flow has a unique fixed point solution
\be
	\mbox{s\GFP{}}^{\rm GS}: \,  \lambda_* = 0.193,  g_* =0.707, 
	\sigma_* = - 0.049,  
\ee 
with stability coefficient $\theta_3 = -2$.  This is the analogue of
the Gau\ss{}ian fixed point for $\sigma$ shifted by the finite
interactions of $g$ and $\lambda$ at the \NGFP{}
\eqref{fpstructure}. The stability coefficient indicates that the new
direction is irrelevant in agreement with
power-counting
arguments.

Taking into account the full non-linear contributions from the $C^3$
term, the cubic \eqref{victory} again has exactly one real root
\be\label{victory2}
\mbox{\NGFP{}}^{\rm GS}: \, \lambda_* = 0.193, \; g_* = 0.707, \; \sigma_* = -0.305 \, , 
\ee
extending the \NGFP{} known from the Einstein-Hilbert projection. The
new stability coefficient $\theta_3 = -79.39$ is again negative, so
that the new direction exhibits an even enhanced irrelevance.  In
fact, the positivity of $c_3$ ensures that $\sigma$ always has a fixed
point for which $C^3$ is an irrelevant perturbation.

\begin{figure}
\includegraphics[width=0.48\textwidth]{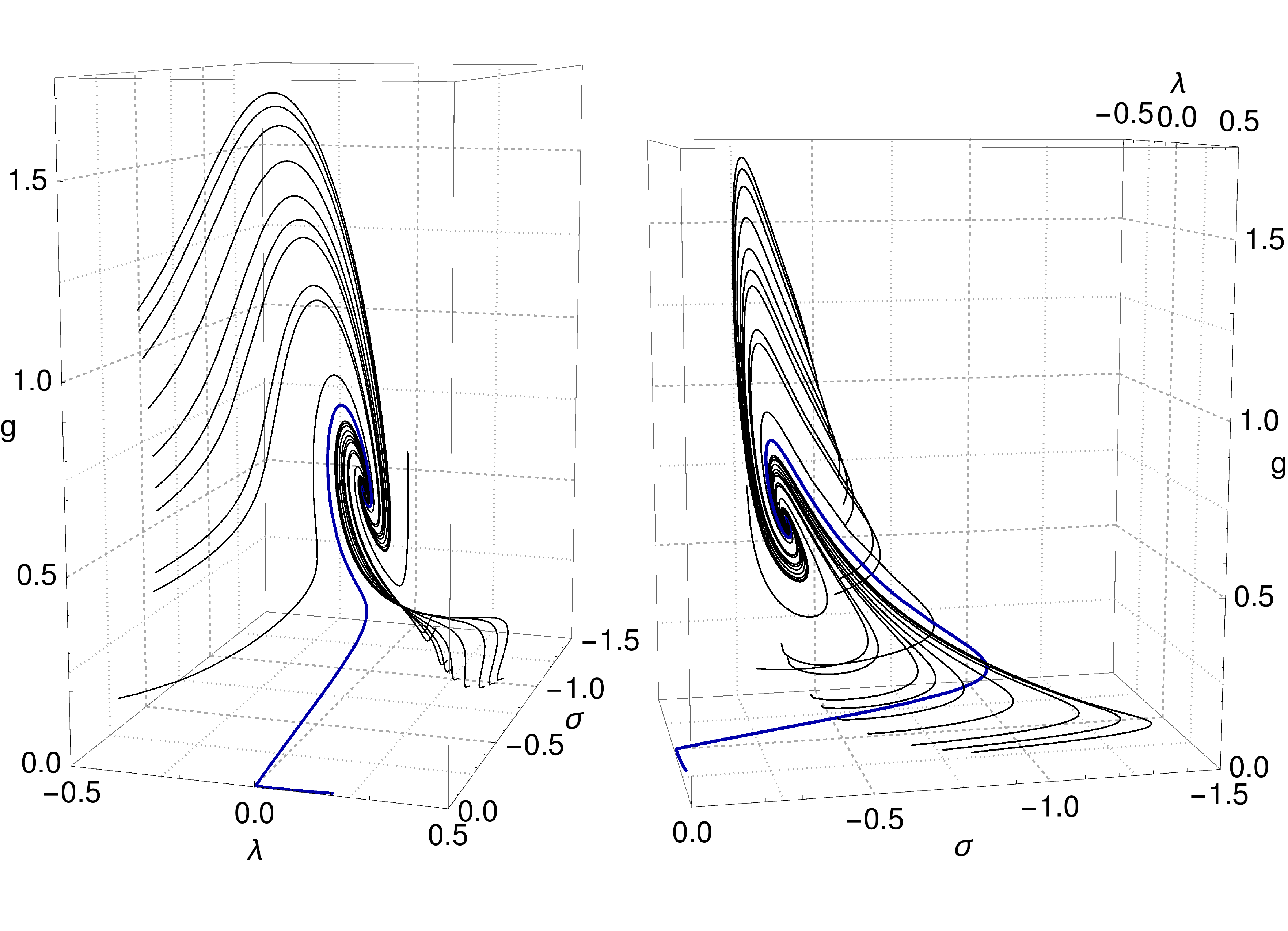}
\caption{Phase diagram in $(g,\lambda,\sigma)$ space from two perspectives depicting trajectories
  emanating from the \NGFP{}. The thick 
line marks a trajectory with
  a long semi-classical regime near the \GFP{}.}
\label{fig:1}
\end{figure}

Fig.~\ref{fig:1} shows the phase diagram in the theory space spanned
by $(g,\lambda,\sigma)$. The flow is governed by the interplay of the
\GFP{} \eqref{gfpgs} and the \NGFP{} \eqref{victory2}. The left panel
depicts a $(g,\lambda)$ perspective illustrating 
that the
inclusion of the Goroff-Sagnotti term leaves asymptotic safety as
observed with the Einstein-Hilbert ansatz
\cite{Reuter:1996cp,Reuter:2001ag} and the $R^2$-extension
\cite{Rechenberger:2012pm} fully intact. The thick 
line
exemplifies a trajectory which crosses over
from the \NGFP{} at
high energies to the \GFP{} at low energies. In the vicinity of the \GFP{}
the trajectory develops a long semi-classical regime where the
couplings scale classically. The right panel presents a $(g,\sigma)$
perspective; following the semi-classical trajectory towards higher
energies, we observe that the Goroff-Sagnotti coupling is first
enhanced but then also attracted by \NGFP{} in the deep \UV. The
dimensionful Goroff-Sagnotti coupling $\bar{\gs}_k \to
\sigma_\ast/k^2$ hence vanishes asymptotically for $k\to\infty$.

While the present ansatz \eqref{ansatz} and calculation scheme give a
unique answer \eqref{victory2} for the fixed point, the number of real
roots of the cubic $\beta$ function \eqref{victory} depends
sensitively on the fixed point values for $g$ and $\lambda$. The
inclusion of higher-order operators thus has the potential to yield
three fixed points. This does, however, not change our conclusion
about the irrelevance of the Goroff-Sagnotti term, as two of these
fixed points have properties equivalent to those discussed above. As
an example, let us consider the case where we neglect the cosmological
constant, setting $\lambda = 0$ at all scales. Then, the \NGFP{} for
Newton's constant has three extensions to the $g, \sigma$-plane. The
one corresponding to \eqref{victory2} is located at $g_*
=12\pi/23\simeq 1.639, \sigma_* = -0.226$ and has stability
coefficients $\theta_1=23/11\simeq 2.09$ and $\theta_3 =
-77.38$. A second fixed point with the same $g_*$ and $\theta_1$
corresponds to the shifted Gau\ss{}ian fixed point for the
Goroff-Sagnotti coupling with $\sigma_* = -0.0023$ and stability
coefficient $\theta_3 = -6.06$.  This confirms the existence of the
\NGFP{} also in the zero-cosmological constant case analyzed by Goroff
and Sagnotti.

\section{Conclusions}
We have studied the non-perturbative renormalization flow of gravity
projected onto the Einstein-Hilbert action supplemented by the
two-loop counterterm found by Goroff and Sagnotti. All versions of the
$\beta$ functions including the Einstein-Hilbert induced
approximation, the zero cosmological-constant limit, and the $\beta$
functions including the full feedback of the counterterm, possess a
non-Gau\ss ian fixed point in agreement with the gravitational
asymptotic safety scenario. This settles a long-standing question
demonstrating that this perturbative counter\-term does not have the
power to destroy the non-Gau\ss ian fixed point seen in the
Einstein-Hilbert projection.  Also the existence of trajectories with
a (semi-)classical low-energy regime is left untouched.  Together with
the recent construction of fixed functionals
\cite{Benedetti:2012dx,Demmel:2012ub,Dietz:2012ic,Dietz:2013sba,Benedetti:2013jk,Demmel:2013myx,Demmel:2014sga,Demmel:2014hla,Bridle:2013sra,Demmel:2015oqa,Dietz:2015owa,Borchardt:2015rxa,Ohta:2015efa,Ohta:2015fcu},
the verification of locality \cite{Christiansen:2015rva}, and first
steps towards clarifying unitarity \cite{Nink:2015lmq}, this
constitutes hard evidence that the asymptotic safety program indeed
can give rise to a consistent quantum theory of gravity within the
framework of quantum field theory along the lines envisioned by
Weinberg \cite{Weinberg:1980gg}.
\medskip

\acknowledgments

We thank M.~Ammon, K.\ Krasnov, A.~Wipf, and O.~Zanusso
for interesting discussions. S.~L.\ and F.~S.\ thank the organizers of
the ``Quantum Gravity in Paris 2015'' workshop for hospitality during
the initial stage of the project. The research of F.~S.\ is supported
by the Netherlands Organisation for Scientific Research (NWO) within
the Foundation for Fundamental Research on Matter (FOM) grants
13PR3137 and 13VP12. H.~G., B.~K.\ and S.~L.\ acknowledge support by
the DFG under grants No. GRK1523/2, Gi328/7-1, and Wi 777/11-1.

\bibliography{general_bib}

\end{document}